\documentstyle[aps,epsfig,float,twocolumn,prl]{revtex}
\newcommand{\be}{\begin{equation}}
\newcommand{\ee}{\end{equation}}

\begin{document}
\twocolumn[\hsize\textwidth\columnwidth\hsize\csname @twocolumnfalse\endcsname
\draft
\title{Eigenvalue Distribution In The Self-Dual Non-Hermitian Ensemble}
\author{M. B. Hastings}
\address{Physics Department, Jadwin Hall\\
Princeton, NJ 08544\\
hastings@feynman.princeton.edu}
\date{16 September 1999}
\maketitle
\begin{abstract}
We consider an ensemble of self-dual matrices with arbitrary complex
entries.  This ensemble is closely related to a previously defined ensemble
of anti-symmetric matrices with arbitrary complex entries.  We study the
two-level correlation functions numerically.  Although no evidence of
non-monotonicity is found in the real space correlation function, 
a definite shoulder is found.  On the analytical
side, we discuss the relationship between this ensemble and the $\beta=4$
two-dimensional one-component plasma, and also argue that this ensemble, 
combined with other ensembles, exhausts the possible universality classes in 
non-hermitian random matrix theory.  This argument is based on combining
the method of hermitization of Feinberg and Zee with Zirnbauer's classification
of ensembles in terms of symmetric spaces.
\end{abstract}
]
\section{Introduction and Classification}
There are ten known universality classes of hermitian random matrices.  
Dyson\cite{dyson}
proposed the existence of three symmetry classes, depending on spin and
the existence of time reversal symmetry.  These give the three classes known
as Gaussian Unitary, Orthogonal, and Symplectic (GUE, GOE, GSE).  Another
three ensembles are the chiral Gaussian ensembles 
(chGUE, chGOE, chGSE)\cite{chiral}.  These ensembles are of relevance to
low energy QCD.  Altland and Zirnbauer introduced four more ensembles
which can appear in superconducting systems\cite{altland}.
Finally, Zirnbauer demonstrated a relationship between the different classes of
random matrix theory and symmetric spaces, and from this argued that the
ten distinct known universality classes exhausted all possible universality
classes\cite{zirn}.

In this section we discuss various universality classes of 
non-Hermitian random matrices, including the ensemble of arbitrary
self-dual matrices, the subject of this paper.  
We mention the concept of weak non-Hermiticity, but do not
consider it further in this paper.  
We argue that the various classes of non-Hermitian matrices, the
self-dual ensemble and four others,
exhaust all possible universality classes.  Finally, possible applications
of the self-dual ensemble are dealt with, including relations with the 
one-component
plasma.  In section II, we further discuss the relationship
with the one-component plasma.  In section III, numerical results for
the self-dual ensemble are discussed, in particular the eigenvalue density
as a function of radius and the two-eigenvalue correlation functions.

Several ensembles of non-Hermitian random matrices are common in the
literature.
Ginibre\cite{ginibre} introduced three classes of such matrices.  One is
an ensemble of matrices with arbitrary complex elements, one an ensemble
with arbitrary real elements, and the third an ensemble with arbitrary
real quaternion elements.  Another ensemble of non-Hermitian matrices is
an ensemble of complex, symmetric matrices\cite{fyod}.  This ensemble
arises particularly in problems of open quantum systems.  This
gives a total of four known universality classes.

For each of these ensembles, there exists a weakly non-Hermitian
version of that ensemble.  This idea of weak non-Hermiticity was
introduced by Fyodorov et al.\cite{weak}.  In this case the anti-Hermitian
part of the matrix is small; we only consider strongly non-Hermitian
matrices in the present paper and do not consider weakly non-Hermitian
matrices, even though they are the most relevant for scattering problems.

The strongly non-Hermitian ensembles can be obtained from
a general three parameter family of non-Hermitian matrices introduced
by Fyodorov et al.\cite{family}.  This family includes parameters measuring
the strength of the real and imaginary, symmetric and anti-symmetric
parts of the matrix.  By adjusting the parameters, one can obtain various
ensembles.  One possiblity, which does not appear to have been considered
much, is an ensemble of anti-symmetric matrices with arbitrary complex
elements.  

Now, let us show that this ensemble is equivalent to
an ensemble of self-dual matrices with arbitrary complex elements; this
is the ensemble considered in this paper.  Let $A$ be an arbitrary
anti-symmetric matrix.  Let $Z$ be the matrix given by
\be
\pmatrix{ 0 & 1 \cr
-1 & 0 \cr
& & 0 & 1 \cr
& & -1 & 0 \cr
& & & & .... \cr}
\ee
Then, $Z^{\rm T}=-Z$ and $Z^2=-1$.  Let $M= ZA$.  It is trival to verify that
$Z M^{\rm T} Z = -M$.  So, $M$ is self-dual.  The advantage of using
self-dual matrices instead of anti-symmetric matrices is that self-dual
matrices have pairs of equal eigenvalues while anti-symmetric matrices
have pairs of opposite eigenvalues; this makes the correlation functions
clearer.  When choosing matrices from the ensemble, we will use Gaussian
weight
\be
\label{gw1}
e^{-\frac{1}{2} {\rm Tr}(M^{\dagger}M)}
\ee

Given these five classes, the 3 ensembles of Ginibre as well as the
ensembles of symmetric non-Hermitian and self-dual non-Hermitian,
let us ask whether all possible universality classes of
strongly non-Hermitian random matrices have been found.  Feinberg and
Zee introduced the method of hermitization for non-Hermitian matrices\cite{zee}.
A similar technique was used by Efetov\cite{efetov}.
The basic idea
is to take a non-Hermitian matrix $M-E$, where $E$ is a complex number,
and form the Hermitian
matrix
\be
\label{v1}
H=\pmatrix{ M_h-E_R & M_a-i E_I \cr M_a^{\dagger}+i E_I & -M_h+E_R}
\ee
where $M_h$ is the Hermitian component of $M$ and $M_a$ is the anti-Hermitian
component of $M$ and $E_R$ and $E_I$ are the real and imaginary components of
$E$.
Equivalently, one can form the Hermitian matrix
\be
\label{v2}
H=\pmatrix{ 0 & M-E \cr M^{\dagger}-\overline E & 0}
\ee
From the zero eigenvalues of $H$, one may extract the zero eigenvalues of
$M-E$.
So, to each universality class of non-Hermitian random matrices, there
corresponds a universality class of Hermitian random matrices.

If we hermitize the three non-Hermitian ensembles introduced by Ginibre,
we obtain the three chiral ensembles (chGUE, chGOE, chGSE).
The relation with the chiral ensembles is most clear
using equation (\ref{v2}), instead of equation (\ref{v1}).
If we
hermitize the ensemble of symmetric, complex matrices we obtain the 
ensemble with symmetry class CI, according to the nomenclature of
Altland and Zirnbauer.  If we hermitize the ensemble of self-dual complex
matrices, we obtain the ensemble with symmetry class DIII.  Here
the relation with the Hermitian ensembles is most clear using
equation (\ref{v1}).  The other
five classes of hermitian random matrices cannot be obtained by
hermitizing a non-Hermitian ensemble: the GOE, GUE, and GSE classes lack
the needed block structure, while the C and D ensembles lack the symmetry
that relates the elements in the upper left and lower right blocks.
This suggests that all possible universality classes of non-Hermitian
matrices have been obtained.

One possible interest in the ensemble of self-dual complex matrices would
be experimental, such as in open-systems with spin orbit scattering.  Another
interest is theoretical, considering the relationship of this ensemble
to the $\beta=4$ one-component plasma in two-dimensions.  Although the
level distribution in the ensemble differs from the distribution of
charges in the plasma, there are some close relations between the
two, discussed more in the next section.

It is known that the ensemble of matrices with arbitrary complex elements
is equivalent to the $\beta=2$ plasma.  The correlation function of the
$\beta=2$ system is monotonic, with Gaussian decay.  From perturbative
calculations\cite{pert}, it has been suggested that, for $\beta>2$, the
two-level 
correlation function becomes non-monotonic, indicating the appearance of
short-range order.  This makes it very interesting to examine the
correlation function of the ensemble of self-dual matrices, although
no significant sign of any non-monotonicity is found here in the
numerical calculations.

Numerical calculations on the one-component plasma\cite{phase} 
suggest that there
is a phase transition at $\beta \approx 144$; so, any order that exists for
$\beta=4$ must be short range.  An exact study for finite number of 
particles\cite{finite}
showed non-monotonicity of the correlation functions for $\beta=4,6$.  Even
for $\beta=4$ there is a definite peak in the correlation function.
\section{$\beta=4$ One-Component Plasma}
Consider a system of $N$ particles, located at positions $z_i$,
with partition function
\be \label{partfn}
\int dz_i d\overline z_i \prod\limits_{i=1}^{N}
e^{-{|z_i|^2}} \prod\limits_{i<j} e^{\beta {\rm log}
(|z_i-z_j|)}
\ee
This defines the two-dimensional one-component plasma.  For $\beta=4$,
there exists some relation between this system and the ensemble considered
here.

First, the density of the plasma, $\rho$, is equal to $\frac{1}{2\pi}$, where
the density is measured in charges per unit area.  The plasma has constant
charge density $\rho$ in a disc about the origin, and vanishing charge
density outside.  The self-dual ensemble has the same charge density, as
found numerically in the next section, and as can be shown with a replica
or SUSY technique.

Second, there exists a relationship between the joint probability
distribution of the eigenvalues of $M$ and the probability distribution
of charges in the one-component plasma.  The j.p.d. of the eigenvalues
of $M$ {\it is different from the charge distribution in the
plasma}, but we will argue that for widely separated eigenvalues the
j.p.d. of the eigenvalues behaves the same as
the probability distribution of the charges.

Let $M$ be a matrix within the ensemble of self-dual, complex matrices.
We can write $M$ as $M=X \Lambda X^{-1}$, where $\Lambda$ is a diagonal
matrix of eigenvalues of $M$.  The eigenvalues of $\Lambda$ exist in pairs,
with $[\Lambda,Z]=0$.
The requirement that $M$ be self-dual is equivalent to the requirement
that $X^{T} Z=Z X^{-1}$ and $Z (X^{-1})^{T}=X Z$; if this constraint on
$X$ holds it is easy to verify that $Z M^{\rm T} Z = -M$.

If we were to impose the additional constraint on $X$ that $X$ be unitary, then
we would find that $X$ must be an element on the symplectic group.
In this case, with $X$ in the symplectic group, the matrix $M$ must be
normal, such that $[M,M^{\dagger}]=0$.
In this case the distribution of eigenvalues of
$M$ exactly matches the charge distribution in the $\beta=4$ plasma.

In the general case, $M$ is not normal and $X$ is not unitary, and 
the distribution of eigenvalues of
$M$ will be different from the charge distribution of the plasma.
Still, consider a situation in which we fix $\Lambda$ and integrate
over $X$, with Gaussian weight $e^{-\frac{1}{2} {\rm Tr}(M^{\dagger}M)}$.  
This is how one obtains the j.p.d. of the eigenvalues.

The measure $[{\rm d}M]$ on matrices $M$ is equivalent to the measure
$[{\rm d}\lambda_i][{\rm d}X]\prod\limits_{i<j} |\lambda_i-\lambda_j|^8$.
The j.p.d. of the eigenvalues is defined by
\be
\prod\limits_{i<j} |\lambda_i-\lambda_j|^8
\int [{\rm d}X] e^{-\frac{1}{2}{\rm Tr}(M^{\dagger}M)}
\ee
with $M=X\Lambda X^{-1}$.
The Gaussian weight, $e^{-\frac{1}{2}{\rm Tr}(M^{\dagger}M)}$,
will depend on $X$.  It will be greatest when $X$ is chosen to be symplectic,
so that $M$ is normal.  If the eigenvalues of $\Lambda$ are well separated,
then the exponential in the Gaussian weight will be large, and we can 
evaluate the integral by a saddle point method: we will restrict
our attention to a saddle point manifold of matrices $M$ which are normal, as
well as weak fluctuations away from this saddle point manifold.
If we parametrize the fluctuations away from the saddle point manifold
and then treat these fluctuations in a Gaussian approximation, valid when the
eigenvalue separation is large, we obtain that the j.p.d. for the self-dual
ensemble is equal to, in this particular approximation,
\be
\prod\limits_{i=1}^{N}
e^{-{|z_i|^2}} \prod\limits_{i<j} 
(|z_i-z_j|)^{4}
\prod\limits_{i=1}^{N} dz_i d\overline z_i 
\ee
up to a constant factors.
This is, of course, the same as the probability distribution of the charges
in the one-component plasma at $\beta=4$.

In general, we expect that for well separated 
eigenvalues, the level repulsion in the self-dual ensemble will match
the charge repulsion in the plasma; it is only the short distance interaction
that will be different.  Further, it may be shown explicitly by calculations
on small matrices that the short distance interactions in the
j.p.d. for the self-dual ensemble
cannot be written as a product of two-body terms.

Given these similarities, one might hope that the correlation functions of
the self-dual ensemble will shed some light on correlations within the
plasma.  In the next section, we discuss a numerical investigation of
the self-dual ensemble.
\section{Numerics}
Mathematica was used to generate
4940 600-by-600 self-dual matrices.  The matrices were chosen with Gaussian
weight
$e^{-\frac{1}{2}{\rm Tr}(M^{\dagger} M)}$
as in equation (\ref{gw1}).
The matrices have 300 pairs of eigenvalues.  A picture of these eigenvalues
for a typical matrix is shown in figure 1.\begin{figure}[!t]
\begin{center}
\leavevmode
\epsfig{figure=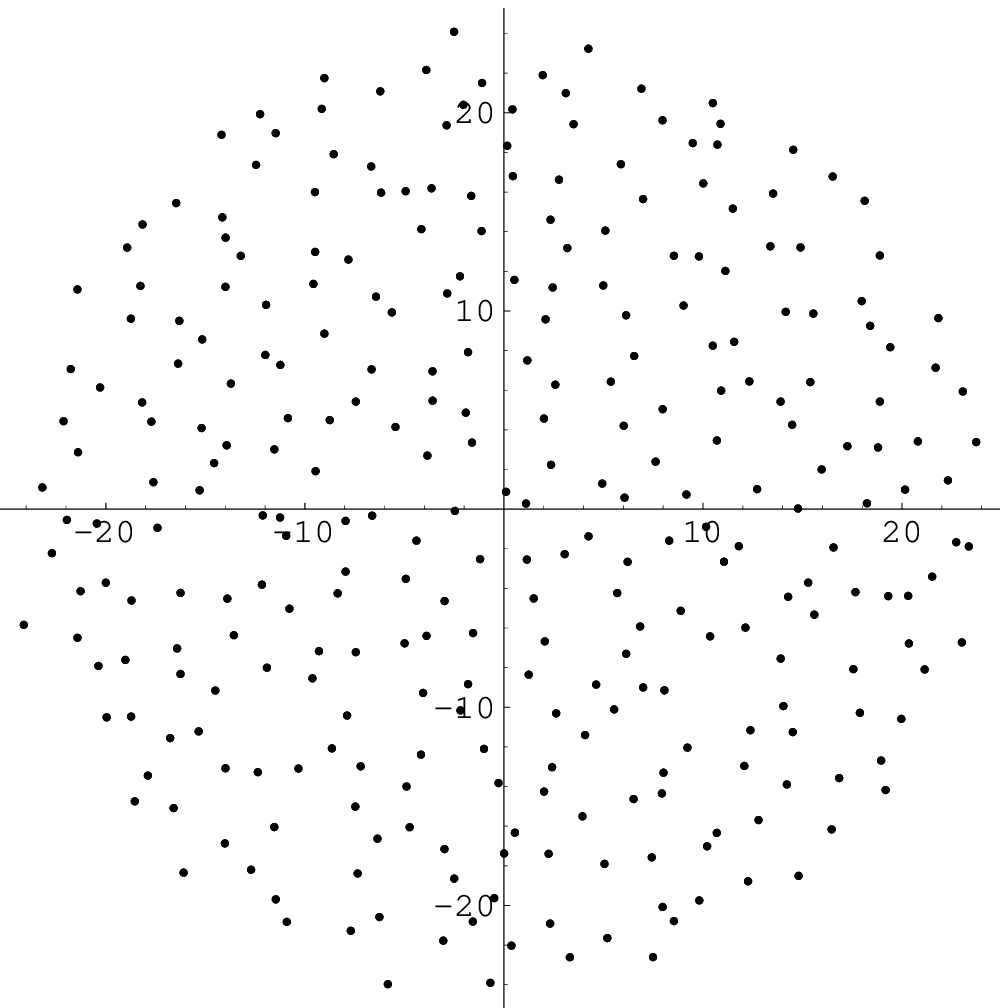,height=8cm,angle=0}
\end{center}
\caption{Plot of eigenvalues for a typical 600-by-600 matrix}
\label{fig1}
\end{figure}

The eigenvalue density as a function of radius is shown in 
figure 2.
The
density obeys the circular law\cite{ginibre,circle}: 
it is nonvanishing and roughly constant
within a disc, and vanishing outside.  For the $\beta=4$ one component
plasma, with a confining potential $e^{-\overline z z}$ (see equation
(\ref{partfn}), the expected density of particles per unit
area, from the circular law, is $\frac{1}{2\pi}$.  The single particle 
eigenvalue density observed numerically for the self-dual
matrices agrees with this result; note that since eigenvalues come in
pairs, then we expect $e^{-\overline z z}$ to be the confining potential
that corresponds to the weight of equation (\ref{gw1}) as each eigenvalue
in the pair contributes a factor of $e^{\frac{\overline z z}{2}}$.
\begin{figure}[!t] 
\begin{center}
\leavevmode
\epsfig{figure=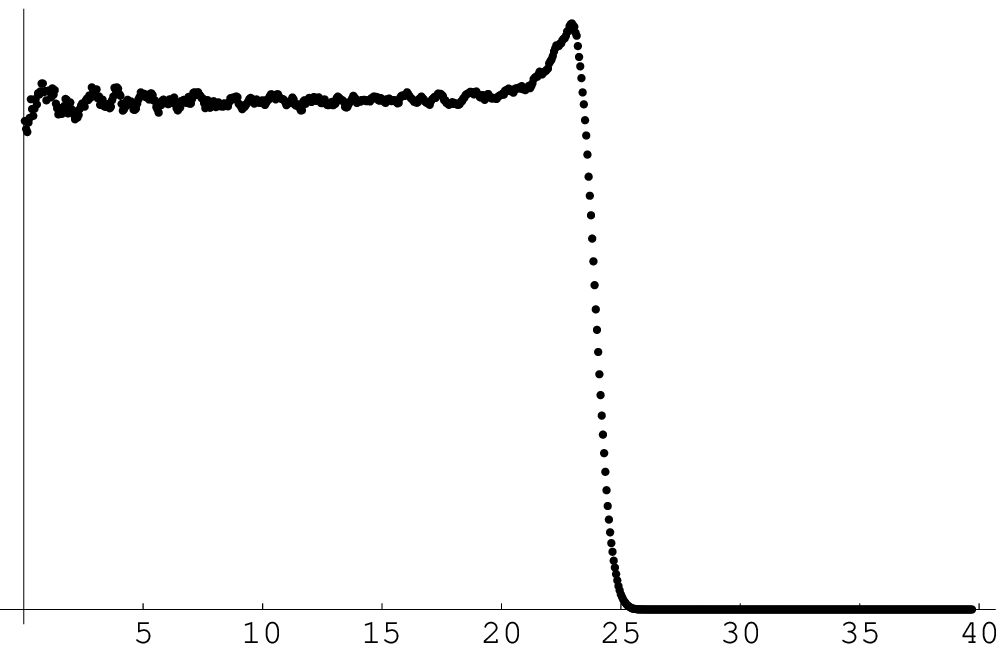,height=5cm,angle=0}
\end{center}
\caption{Average eigenvalue density as a function of radius}
\label{fig2}
\end{figure}

One interesting feature of figure 2 is that the eigenvalue density near the edge
rises before dropping.  It is not clear why this happens.

The two-level correlation function is shown in figures 3 and 4.  In figure
3, we look at all eigenvalues within a distance of 6 or less from the origin, 
and plot the probability to find another eigenvalue at given distance from the
first eigenvalue. In figure 4, to reduce effects due to the finite size
of the matrix $M$, we require that the first eigenvalue lie within a
distance of 3.5 or less from the origin.  No significant differences are
found between figure 3 and figure 4, indicating that 
the effects due to the finite
size of $M$ are small even in figure 3.
\begin{figure}[!t]
\begin{center}
\leavevmode
\epsfig{figure=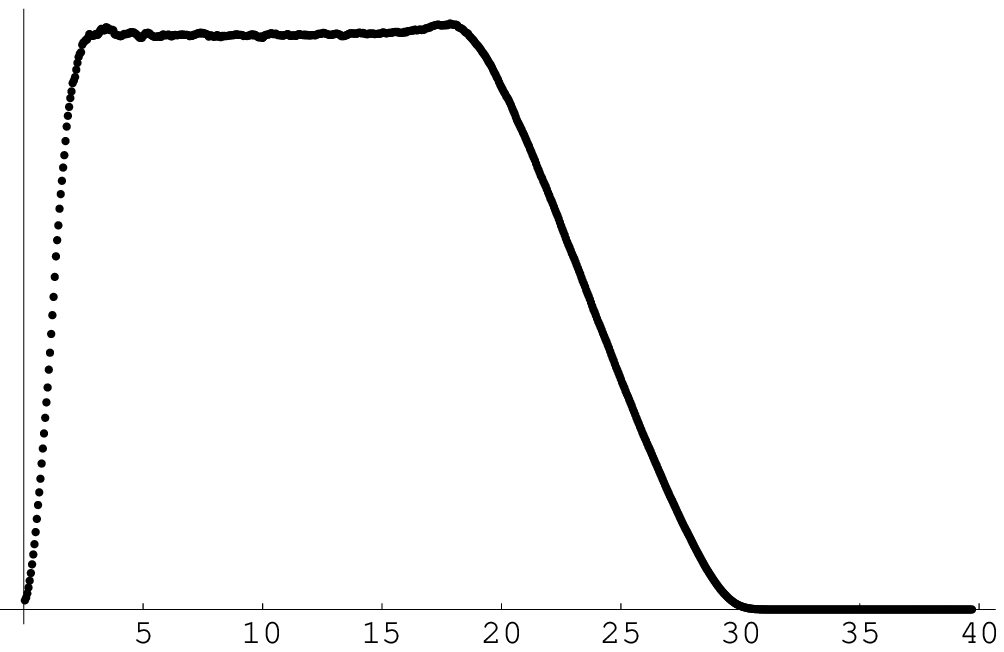,height=6cm,angle=0}
\end{center}
\caption{Average two level correlation function.  See text.}
\label{fig3}
\end{figure}
\begin{figure}[!t]
\begin{center}
\leavevmode
\epsfig{figure=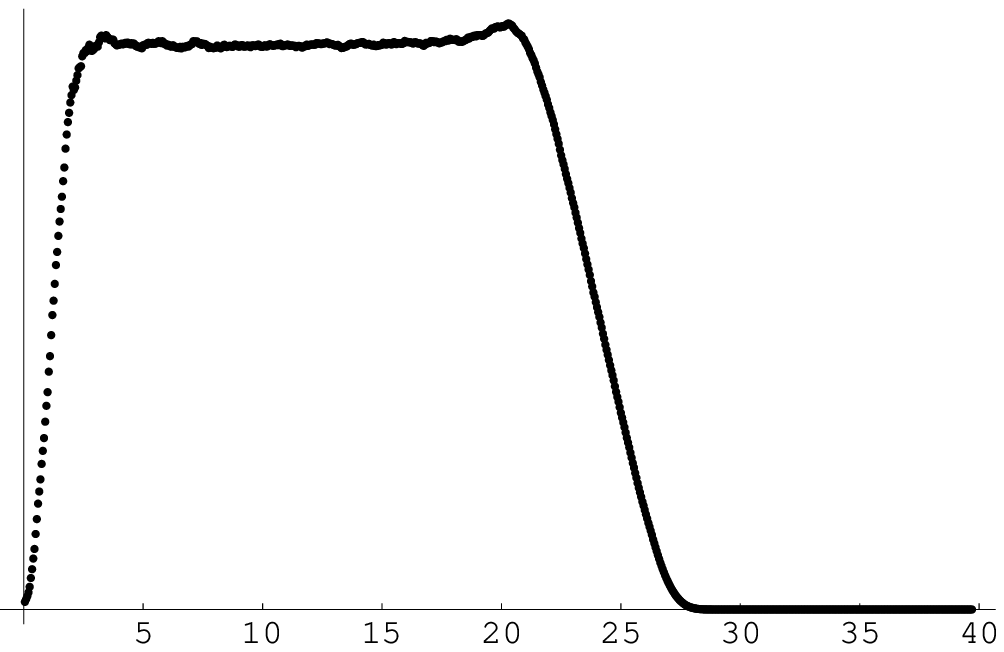,height=6cm,angle=0}
\end{center}
\caption{Average two level correlation function.  See text.}
\label{fig4}
\end{figure}

One can see finite size effects in both figure 3 and 4 for large
distances.  The correlation function rises for distances of around 20.
This is simply due to the rise in eigenvalue density near the edge,
as shown in figure 2, and has no deep meaning.

Looking at figures 3 and 4, there is a definite ``shoulder" at a distance
of slightly less than 3.  
There is no definite sign of any non-monotonicity; certainly,
if there is any peak in the correlation function near the shoulder, it
is much smaller than the peak found in the
$\beta=4$ plasma\cite{finite}.  As a quick estimate of the expected
spacing between levels, assume that the levels formed a perfect
hexagonal lattice, so that they are very ordered, and packed as closely as
possible.  In this case, if the levels have a density of $\frac{1}{2\pi}$,
then the closest spacing between levels is $\frac{2 \sqrt{\pi}}{3^{1/4}}$,
which is approximately 2.7.  For other arrangements of levels, the
spacing will be slightly less.
This length agrees quite well with the size of the shoulder.
So, the shoulder length matches reasonably with the length scale
expected from the particle spacing.
\section{Conclusion}
In conclusion, we have considered an ensemble of strongly non-Hermitian,
self-dual matrices.  The two-level correlation function of this ensemble
is particularly interesting, although the hoped for non-monotonicity
has not emerged.  It seems that all possible universality classes of
non-Hermitian matrices are now known.


\begin{thebibliography}{99}
\bibitem{dyson}
F. J. Dyson, {\it J. Math. Phys.} {\bf 3}, 140 (1962).

\bibitem{chiral}
J. J. M. Verbaarschot and I. Zahed, {\it Phys. Rev. Lett.}
 {\bf 70}, 3853 (1993);
J. J. M. Verbaarschot, {\it Phys. Rev. Lett.} {\bf 72}, 2531 (1994).

\bibitem{altland}
A. Altland and M. R. Zirnbauer, {\it Phys. Rev. B} {\bf 55}, 1142 (1997).

\bibitem{zirn}
M. R. Zirnbauer, {\it J. Math Phys} {\bf 37}, 4986 (1996).

\bibitem{ginibre}
J. Ginibre, {\it J. Math. Phys.} {\bf 6}, 440 (1965).

\bibitem{fyod}
H.-J. Sommers, Y. V. Fyodorov, and M. Titov, preprint chao-dyn/9807015;
Y. V. Fyodorov, H.-J. Sommers, {\it J. Math. Phys.} {\bf 38}, 1918 (1997).

\bibitem{family} 
Y. V. Fyodorov, B. Khoruzhenko, and H.-J.  Sommers, 
{\it Ann. Inst. H. Poincare, Phys. Theor.}, {\bf 69}, 449 (1998).

\bibitem{weak} 
Y. V. Fyodorov, B. Khoruzhenko, and H.-J.  Sommers, 
{\it Phys. Lett. A}, {\bf 226}, 46 (1997);
Y. V. Fyodorov, B. Khoruzhenko, and H.-J.  Sommers, 
{\it Phys. Lett. A}, {\bf 79}, 557 (1997).

\bibitem{zee}
J. Feinberg and A. Zee, preprint cond-mat/9703087.

\bibitem{efetov}
K. B. Efetov, preprint cond-mat/9702091.

\bibitem{pert}
B. Jancovici, {\it Phys. Rev. Lett.} {\bf 46}, 386 (1981).

\bibitem{phase}
J. M. Caillol et. al., J. Stat. Phys {\bf 28} 325, (1982).

\bibitem{finite}
G. Tellez, P. J. Forrester, preprint cond-mat/9904388.

\bibitem{circle}
L. Girko, Theor. Probab. Appl. {\bf 29}, 694 (1985); {\bf 30}, 677 (1986).

\end{thebibliography}
\end{document}